\documentclass[preprint2]{aastex6}

\fullcollaborationName{}

\shorttitle{Imaging the Multi-Gapped Disk of HD 169142}
\shortauthors{Mac\'{\i}as et al.}

\begin{document}

\title{Imaging a Central Ionized Component, a Narrow Ring, and the CO Snowline in the Multi-Gapped Disk of HD 169142}

\author{Enrique Mac\'{\i}as}
\affil{Department of Astronomy, Boston University, 725 Commonwealth Avenue, Boston, MA 02215, USA \\
Instituto de Astrof\'\i sica de Andaluc\'\i a (CSIC) Glorieta de la Astronom\'\i a s/n E-18008 Granada, Spain
{\tt emacias@bu.edu}}

\author{Guillem Anglada, Mayra Osorio}
\affil{Instituto de Astrof\'\i sica de Andaluc\'\i a (CSIC) Glorieta de la Astronom\'\i a s/n E-18008 Granada, Spain}

\author{Jos\'e M. Torrelles \thanks{The ICC (UB) is a CSIC-Associated Unit through the ICE}}
\affil{Institut de Ci\`encies de l'Espai (CSIC) and Institut de 
Ci\`encies del Cosmos (UB)/IEEC, Can Magrans S/N, Cerdanyola del Vall\`es (Barcelona), Spain}

\author{Carlos Carrasco-Gonz\'alez}
\affil{Instituto de Radioastronom\'{\i}a y Astrof\'{\i}sica 
UNAM, Apartado Postal 3-72 (Xangari), 58089 Morelia, Michoac\'an, Mexico}

\author{Jos\'e  F. G\'omez}
\affil{Instituto de Astrof\'\i sica de Andaluc\'\i a (CSIC) Glorieta de la Astronom\'\i a s/n E-18008 Granada, Spain}

\author{Luis F. Rodr\'{\i}guez \& Anibal Sierra}
\affil{Instituto de Radioastronom\'{\i}a y Astrof\'{\i}sica 
UNAM, Apartado Postal 3-72 (Xangari), 58089 Morelia, Michoac\'an, Mexico}

\begin{abstract}
We report Very Large Array observations at 7 mm, 9 mm, and 3 cm toward the
pre-transitional disk of the Herbig Ae star HD 169142. These observations have
allowed us to study the mm emission of this disk with the highest angular
resolution so far ($0\rlap.''12\times0\rlap.''09$, or 14 au$\times$11 au, at 7
mm). Our 7 and 9 mm images show a narrow ring of emission at a radius of $\sim25$
au tracing the outer edge of the inner gap. This ring presents an asymmetric morphology that could be produced by dynamical interactions between the disk and forming planets. Additionally, the azimuthally averaged radial intensity profiles of the 7 and 9 mm images confirm the presence of the previously reported gap at $\sim45$ au, and reveal a new gap at $\sim85$ au. We analyzed archival DCO$^+$(3-2) and C$^{18}$O(2-1) ALMA observations, showing that the CO snowline is located very close to this third outer gap. This suggests that growth and accumulation of large dust grains close to the CO snowline could be the mechanism responsible for this proposed outer gap. Finally, a compact source of emission is detected at 7 mm, 9 mm, and 3 cm toward the center of the disk. Its flux density and spectral index indicate that it is dominated by free-free emission from ionized gas, which could be associated with either the photoionization of the inner disk, an independent object, or an ionized jet. 
\end{abstract}


\section{Introduction} \label{sec:intro}

Planetary systems are formed in circumstellar disks around pre-main sequence stars. Tidal interactions between the forming planets and the disk can result in complex substructures such as cavities, gaps, spirals, or lopsided rings \citep{bar14}. Studying disks showing these features could provide us with critical information about the planetary formation process itself. In particular, transitional disks, which are protoplanetary disks with central dust gaps or cavities typically of tens of au in size \citep{str89}, appear as excellent candidates to study the first stages of planetary formation.

Cavities in transitional disks were first identified through modeling of their spectral energy distributions (SEDs) \citep{cal05}. This modeling also lead to the discovery of a subfamily of transitional disks, the so-called pre-transitional disks, which are thought to present a residual inner disk inside the cavity that can still emit significantly at near-IR wavelengths \citep{esp07}. (Sub-)mm and polarimetric IR observations have been able to image several of these disks and confirm the presence of the central cavities or gaps (e.g., \citealp{and11,qua11}). Since their discovery, a number of mechanisms have been proposed to explain these inner clearings of dust (\citealp{esp14} and references therein). Nevertheless, observations seem to indicate that most cavities in transitional disks are created by dynamical interactions with orbiting substellar or planetary companions \citep{and11,esp14}. 

Until recent years, (sub-)mm observations lacked the sensitivity and angular resolution necessary to reach distances very close to the central star.
With the outstanding angular resolution provided by the most extended baselines of the Atacama Large Millimeter/Submillimeter Array (ALMA), as well as with the new capabilities of the Karl G. Jansky Very Large Array (VLA), it is now possible to attempt this type of studies. In particular, recent ALMA observations have revealed the presence of very small central cavities, few au in size, in some transitional disks (e.g., XZ Tau B: \citealp{oso16}; TW Hya: \citealp{and16}). On the other hand, in a few cases it has been possible to detect compact central emission inside the cavity of transitional disks \citep{ise14,rod14,and16}. 
This central emission has been associated with an inner disk (emission either from dust or from photoionized gas) or with an ionized jet.

Additionally, (sub-)mm ALMA observations have also revealed the presence of several gaps and rings up to distances of $\sim90$ au from the star in the protoplanetary disk around HL Tau \citep{alm15}. These gaps, however, might have a different origin to those observed at the inner regions of transitional disks. The young age of HL Tau, as well as the fact that some gaps are very narrow and appear at very long distances from the star, have lead to question whether a planet could produce this type of gaps. This has resulted in a number of studies proposing new physical processes that could create similar gap structures -- e.g. zonal flows in magnetized disks \citep{bai14}, magneto-rotational instability at the dead zone outer edge \citep{flo15}, sintering-induced gaps and rings \citep{oku16}, or grain growth close to snowlines in the disk \citep{ros13,zha15}. Gaps produced by these mechanisms could also be present in older protoplanetary and transitional disks, but they would have remained unnoticed in previous (sub-)mm observations due to a lack of sensitivity and angular resolution. In fact, recent ALMA observations, with a similar angular resolution to the HL Tau observations, have revealed the presence of similar ringed substructures in the transitional disk of TW Hya \citep{and16} and in the protoplanetary disk of HD 163296 \citep{ise16}. Similar observations of protoplanetary disks at these high angular resolutions could show whether multi-gap structures are ubiquitous in protoplanetary disks, and will help to identify the physical mechanisms responsible for their creation.

HD 169142 is a Herbig Ae star ($M_{\star}\simeq1.65$--$2$ M$_{\odot}$, age $\simeq5$--$11$ Myr, \citealp{blo06,man06}) surrounded by an almost face-on ($i\simeq13^{\circ}$; \citealp{ram06}) pre-transitional disk (\citealp{oso14} and references therein). \citet{oso14} presented VLA observations at 7 mm toward HD 169142, detecting a bright ring of dust emission at a radius of $\sim0\rlap.''2$ surrounding a central cavity, as well as an outer gap from $\sim0\rlap.''28$ to $\sim0\rlap.''48$, coincident with the results from IR polarized scattered light images \citep{qua13,mom15,mon17}. These results have been confirmed by recent ALMA 1.3 mm continuum and CO observations, showing that the two dust gaps are filled in with gas with a significantly reduced density at radii smaller than $\sim0\rlap.''48$ \citep{fed17}. The 7 mm emission ring imaged by \citet{oso14} shows an azimuthally asymmetric morphology, reminiscent of the lopsided morphology produced as a consequence of dust trapping in planet-induced vortices \citep{bir13}. 
These authors also reproduced the SED and 7 mm radial intensity profile of HD 169142 with a disk model that included the central cavity and the outer gap. Their results show that an inner residual disk is required to fit the SED and concluded that the disk of HD 169142 is a pre-transitional disk with two gaps. Their results also suggested that planet formation is the most likely origin for both detected gaps. In fact, a substellar or planetary companion candidate has been detected within the inner gap of the disk \citep{reg14,bil14}, which supports a planet-induced origin for this gap. 

We note that most of the previous studies of HD 169142 adopted a distance of 145 pc \citep{van05}. However, the recent publication of the first data release of Gaia has revealed that the distance to HD 169142 is $117\pm4$ pc \citep{gai16}. This represents a decrease of $\sim20\%$ from the value of 145 pc adopted in the literature. The same reduction has been therefore applied throughout this article to the sizes of the disk structures measured here and in the literature. 

In this paper we present new high angular resolution VLA observations at 7 mm, 9 mm, and 3 cm toward the pre-transitional disk around HD 169142, revealing the presence of a new third gap in the disk as well as a free-free thermal emission source inside the inner gap.

\section{Observations} \label{sec:observations}

We performed observations using the VLA of the National Radio Astronomy Observatory (NRAO)\footnote{The NRAO is a facility of the National Science Foundation operated under cooperative agreement by Associated Universities, Inc.} in the A and BnA configurations at Q ($\sim$7 mm), Ka ($\sim$9 mm), and X ($\sim$3 cm) bands. Archival observations at K ($\sim$1.3 cm; C and DnC configurations) and C bands ($\sim$5 cm; A configuration) were also used (see Table \ref{Tab:obs}). Amplitude calibration was performed by observing 3C286, with an expected uncertainty in the flux scale of $\sim10\%$. 3C286 was also used as the bandpass and delay calibrator, whereas J1820-2528 was used as the complex gain calibrator.

\begin{deluxetable*}{lccccccc}
\tablecaption{VLA observations \label{Tab:obs}}
\tablehead{
\colhead{} & \colhead{Central} & \colhead{} &\colhead{Array} & \colhead{Observation} & \colhead{Project} & \colhead{On-source}\\
\colhead{Band} & \colhead{Frequency} & \colhead{Bandwidth} & \colhead{Configuration} & \colhead{Date} & \colhead{Code} & \colhead{time}\\
\colhead{} & \colhead{(GHz)} & \colhead{(GHz)} & \colhead{} & \colhead{} & \colhead{} & \colhead{(min)}\\
}
\startdata
Q  & 44 & 8 &  A    & 2014-Mar-06 & 14A-496 & 63.9  \\
   &    &   &       & 2014-Mar-04 &         & 63.9  \\
   &    &   &       & 2014-Feb-27 &         & 63.9  \\
Q  & 44 & 8 &  BnA  & 2014-Jan-25 & 13B-260 & 72.0  \\
Q  & 44 & 8 &   B   & 2013-Sep-28\tablenotemark{a} & 13B-260 & 72.0  \\
Ka & 33 & 8 &  A    & 2014-Feb-25 & 14A-496 & 45.4  \\
K  & 21 & 2 &  C    & 2012-Jan-29 &  AC982  &  6.0  \\
   &    &   & DnC   & 2010-Sep-30 &         &  6.0  \\
X  & 10 & 4 &  A    & 2014-Mar-06 & 14A-496 & 15.3  \\
   &    &   &       & 2014-Mar-04 &         & 15.3  \\
   &    &   &       & 2014-Feb-27 &         & 46.0  \\
   &    &   &       & 2014-Feb-25 &         & 15.3  \\
X  &  9 & 2 &  BnA  & 2014-Jan-25 & 13B-260 &  7.3  \\
X  &  9 & 2 &   B   & 2013-Sep-28\tablenotemark{a} & 13B-260 &  7.3  \\
C  &5.5 & 2 &   A   & 2012-Dec-23 & 12A-439 &  8.0  \\
   &    &   &       & 2012-Nov-12\tablenotemark{a} &         &  8.0  \\
\enddata
\tablenotetext{a}{~Data reported in \citet{oso14}.}
\end{deluxetable*}

The observations were reduced and calibrated with the reduction package Common Astronomy Software Applications (CASA; version 4.5.3; \citealp{mcm07})\footnote{https://science.nrao.edu/facilities/vla/data-processing}. Each data set was processed through the VLA calibration pipeline integrated within CASA. After each run of the pipeline the calibrated data were inspected. Then, we performed additional data flagging and re-ran the pipeline as many times as needed. 

Deconvolved images were produced with the CLEAN task of CASA. A multi-scale multi-frequency deconvolution algorithm was used to take into account the frequency dependence of the emission within each band \citep{rau11}. Data from each observing session were first imaged independently to check for possible errors in the absolute position, without finding any significant shift. Then, for each band, we combined the data from the different epochs and configurations in order to obtain higher sensitivity images. A {\it uvtaper} was applied to the Ka band visibilities in order to improve the signal to noise ratio of the image.

In addition, we report unpublished archival Atacama Large Millimeter/submillimeter Array (ALMA) data of the DCO$^+$(3-2) transition (rest frequency 216.112 GHz), and we reanalyze data of the C$^{18}$O(2-1) transition (rest frequency 219.560 GHz). The observations were carried out on 2015 August 30 (project code: 2013.1.00592.S) and are described in \citet{fed17}, where the C$^{18}$O(2-1) data were first reported. Data calibration was performed using the ALMA pipeline within CASA (version 4.3.1). Deconvolved images were then obtained using the CLEAN task with natural weighting. In addition, a {\it uvtaper} was applied to the DCO$^+$(3-2) visibilities in order to achieve a higher signal-to-noise ratio. The rms noise of the C$^{18}$O(2-1) observations is $\sim6$ mJy beam$^{-1}$ (synthesized beam of $0\rlap.''35\times0\rlap.''23$, PA=-74$^{\circ}$) for a channel width of $\sim0.25$ km s$^{-1}$, whereas the rms noise of the DCO$^+$(3-2) observations is $\sim7$ mJy beam$^{-1}$ (synthesized beam of $0\rlap.''50\times0\rlap.''48$, PA=-65$^{\circ}$) for a channel width of $\sim0.25$ km s$^{-1}$.

\section{Results and discussion} \label{sec:results}

\begin{figure*}
\figurenum{1}
\plotone{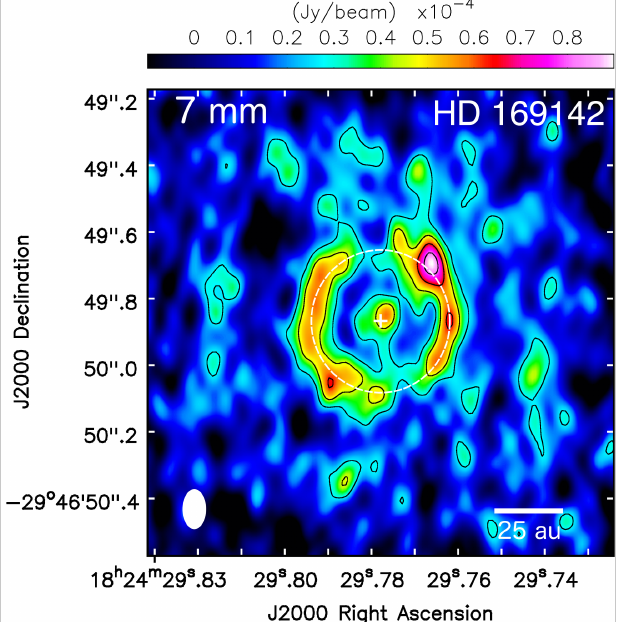}
\label{fig:maps}
\caption{ Natural-weighted VLA image at 7 mm of the transitional disk around HD 169142 (synthesized beam=$0\rlap.''12\times0\rlap.''07$, PA=$0^{\circ}$; shown in the lower-left corner). Contour levels are $-$3, 3, 5, 7, and 9 times the rms of the map, $9.0~\mu$Jy beam$^{-1}$. The dashed ellipse indicates our fit to the ring image at 7 mm. The white cross shows the position of the center of the ring.}
\end{figure*}

A natural-weighted image of the 7 mm emission of HD 169142, obtained from the combination of A, BnA, and B configuration data, is shown in Figure \ref{fig:maps}. The image shows a narrow ring of emission of radius $\sim0\rlap.''21$ ($\sim25$ au at 117 pc) with significant substructure. In addition, the image shows a hint of a second ring of emission at $\sim0\rlap.''50$ ($\sim59$ au) tracing the rim of the second gap detected by \citet{oso14}. A compact emission component is detected inside the inner ring, with its peak of emission displaced a projected distance of $\sim0\rlap.''023$ ($\sim2.7$ au) from its center. In the images made from a single configuration, this central component of emission is only detected in the A configuration data, which have enough angular resolution and sensitivity to separate it from the ring of emission. We do not expect significant proper motions within the time span of the A configuration observations (one week). Thus, we expect that our images obtained combining the A, BnA, and B configuration data will not be affected by these proper motions. The total flux density of the 7 mm emission is $2.0\pm0.4$ mJy, which is consistent with previous measurements \citep{oso14}. The flux density of the central component is $74\pm15~\mu$Jy.

Our new A configuration data, with higher sensitivity and angular resolution, do not confirm the knot of 7 mm emission located $\sim$0.34$''$ ($\sim$40 au) to the south of the central position, suggestive of a protoplanet candidate inside the second gap, that was observed in the \citet{oso14} images. Also, we do not identify radio emission associated with the IR source detected by \citet{reg14} and \citet{bil14} at radius $0\rlap.''16$ and PA = $7^{\circ}$. 

\begin{figure*}
\figurenum{2}
\plottwo{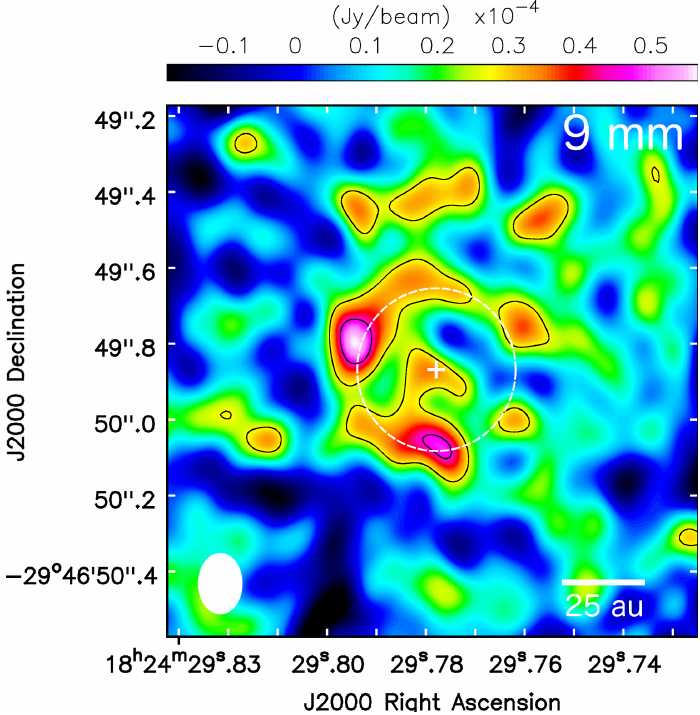}{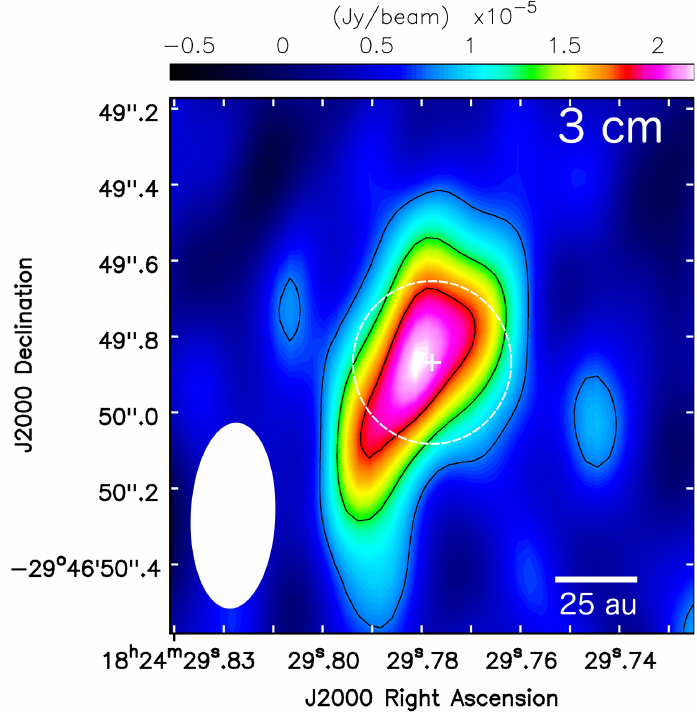}
\label{fig:maps2}
\caption{Left panel: Natural-weighted VLA image of the 9 mm emission (synthesized beam=$0\rlap.''16\times0\rlap.''12$, PA=$0^{\circ}$; shown in the lower-left corner). Contour levels are $-$3, 3, 5, and 7 times the rms of the map, $9.0~\mu$Jy beam$^{-1}$. Right panel: Natural-weighted VLA image of the 3 cm emission (synthesized beam=$0\rlap.''49\times0\rlap.''22$, PA=$-2.5^{\circ}$; shown in the lower-left corner). Contour levels are $-$3, 3, 5, and 7 times the rms of the map, $2.6~\mu$Jy beam$^{-1}$. In both panels the dashed ellipse indicates our fit to the ring image at 7 mm as shown in Figure \ref{fig:maps}. The white cross shows the position of the center of the ring.}
\end{figure*}

The left panel in Figure \ref{fig:maps2} presents the 9 mm image of HD 169142. This image shows a similar morphology to the 7 mm image: a ring of emission with a central radio source inside its cavity. The ring of emission also seems to show some substructure, although the low signal to noise ratio makes it difficult to determine whether this substructure in the 9 mm image is real or due to rms fluctuations. The total flux density of the 9 mm emission is $850\pm150~\mu$Jy, while the flux density of the central emission inside the inner ring is $45\pm14~\mu$Jy. 

The right panel in Figure \ref{fig:maps2} shows an image of the 3 cm emission of HD 169142, obtained by combining the A, BnA, and B configuration observations using natural weighting. The emission is only marginally resolved, with its peak of emission located inside the inner dust gap, very close to the central star. The total flux density at 3 cm is $50\pm10~\mu$Jy. Due to the lower angular resolution of the 3 cm observations, we cannot directly separate in our natural-weighted image the emission of the extended disk from the central radio source. By using a higher weight for the most extended visibilities, we can filter out the extended disk emission and estimate the flux density of the compact central radio source. We used Briggs weighting with a robust parameter of 0.5 (as defined in task CLEAN of CASA) and estimated a flux density at 3 cm of $\sim20\pm5~\mu$Jy for the central radio source. The remaining emission in the natural-weighted image at 3 cm is consistent with the dust flux density of $\sim30~\mu$Jy predicted by the model of \citet{oso14}. 

Finally, no emission was detected in the K and C band observations, with $3\sigma$ upper limits of $420~\mu$Jy beam$^{-1}$ and $27~\mu$Jy beam$^{-1}$, respectively. The C band image was obtained by combining our new data with those previously reported by \citet{oso14} in order to obtain a tighter upper limit. Both limits are consistent with the model presented by \citet{oso14}.

\subsection{Narrow ring}\label{subsec:ring}

As can be seen in Figure \ref{fig:maps}, our 7 mm image shows a narrow ring of emission with a radius of $\sim25$ au. For the width of the ring we estimate a deconvolved FWHM of $\sim8$ au, measured along the E-W direction, where the synthesized beam size is smaller. The ring in our images coincides quite well with the ring imaged in previous near-IR \citep{qua13,mon17}, 7 mm \citep{oso14}, and 1.3 mm \citep{fed17} observations, but our images reveal additional details. 

We have estimated the position of the center of the ring by fitting an ellipse with its major axis along the position angle of the disk, as estimated from previous molecular observations (PA$=5^{\circ}$; \citealp{ram06}), to the ring image, excluding the regions where two dips and a knot of emission are found (see below). From this fit we estimate that the center of the ring in the 7 mm image (epoch $\sim2014.17$) is located at $\alpha({\rm J2000})=18^{\rm h} 24^{\rm m} 29\rlap.^{\rm s}77798\pm0\rlap.^{\rm s}00018$, $\delta({\rm J2000})=-29^{\circ}46'49\rlap.''8673\pm0\rlap.''0028$, which coincides within $\sim4\pm3$ mas with the position of the star HD 169142 given in the Gaia catalog, after correcting for proper motions \citep{gai16}. This result indicates that the ring is very well centered on the star, and that the alignment of the data to make the final image, as well as the quality of its astrometry, are very good. 
From our fit we also estimate a radius of the ring of $0\rlap.''214\pm0\rlap.''004$ ($25.0\pm0.5$ au), which is slightly larger than the radius of $\sim0\rlap.''19$ ($\sim22$ au) estimated from the near-IR scattered light images (\citealp{qua13}). This suggests that the scattered light emission arises mainly from the inner rim or wall of the ring, whereas the 7 mm emission traces mainly the surface density of the large dust grains that peaks at slightly larger radii.

The intensity of the ring in our 7 mm image is significantly asymmetric in azimuth, showing a knot of emission $\sim4~\sigma$ above the average intensity of the ring at PA$\simeq-40^{\circ}$, in agreement with the previous results of \citet{oso14}, who noted this azimuthal asymmetry from 7 mm data of lower angular resolution. We think that this knot represents a real azimuthal asymmetry since an intensity enhancement appears both in the images made from the new A configuration data alone as well as in the lower angular resolution maps reported in \citet{oso14}. The ALMA 1.3 mm images also show a hint of this asymmetry \citep{fed17}, although at a lower significance, probably due to the higher optical depth of the ring at shorter wavelengths, or an accumulation of the large dust grains preferentially traced at 7 mm.

On the other hand, our new 7 mm image also shows significant decreases of intensity or dips in the ring at PA$\simeq0^{\circ}$ and PA$\simeq-170^{\circ}$. As shown by \citet{oso14}, an elongated beam can produce depressions of emission along the direction of the major axis of the beam in an axisymmetric ring. Additionally, noise fluctuations can produce spurious clumpy structures during the deconvolution process. In order to check whether the substructure detected in our images is real or due to these spurious effects, we obtained simulated images with random thermal noise using the model presented by \citet{oso14} and the SIMOBSERVE task in CASA. These simulations showed that, for certain noise structures, intensity depressions similar to the ones in our observations could be formed in the narrow ring, suggesting that the observed dips in our 7 mm observations could be produced because of the combined effect of noise fluctuations and the elongated beam. Nevertheless, none of our simulations produced a bright knot similiar to the one detected in our 7 mm observations, indicating that it is tracing a real azimuthal asymmetry in the ring.

Polarized scattered light images at {\it H} band show an axisymmetric ring with only a possible dip at a PA$\simeq80^{\circ}$ \citep{qua13}, where no significant drop of emission is seen in our images. The fact that the knot of emission in the ring is not present at near-IR wavelengths indicates that it is probably produced by azimuthal asymmetries in the density near the disk mid-plane, to which our 7 mm images are more sensitive, without affecting significantly the distribution of small dust grains in the disk atmosphere, which are traced by the scattered light images.

Azimuthal asymmetries in the large dust grains distribution are expected to be produced by tidal interactions between a forming planet and the disk (e.g., \citealp{bar14}). Hydrodynamic simulations show that planets can create relatively large cavities with vortices at their outer edges. These vortices, in turn, are able to trap the large dust grains in their pressure maxima, producing lopsided asymmetries at mm wavelengths \citep{bir13,zhu14}.
Thus, the interaction between the disk of HD 169142 and the possible forming planets at the inner \citep{reg14,bil14} and second gaps could be responsible for the observed non-axisymmetric structure.

\subsection{Central compact radio source}\label{subsec:central}

Our 7 mm, 9 mm, and 3 cm images have revealed the presence of compact emission inside the bright emission ring, originating near its center (Figs. \ref{fig:maps} and \ref{fig:maps2}). The emission at 7 mm is slightly extended along the E-W direction, with its intensity peak at a projected distance of $\sim3\pm1$ au ($\sim0\rlap.''025\pm0\rlap.''010$) from the central star toward the W direction. The quoted uncertainty is probably only a lower limit, corresponding to the formal error in the position of the emission peak relative to the center of the ring, estimated as $\sim0.5(\theta/\rm{SNR})$ \citep{rei88}, assuming an unresolved source ($\theta=0\rlap.''1$) and a signal-to-noise ratio (SNR) of 5. Further observations are needed to confirm the reality and origin of this possible displacement (see below).

The insufficient sensitivity and angular resolution of the images at 9 mm and 3 cm, respectively, make it difficult to estimate the morphology of the central emission observed at these wavelengths. This central radio source could not be detected in previous observations by \citet{oso14} toward HD 169142 at 7 mm due to the lack of angular resolution. We do not identify radio emission associated with the IR source detected by \citet{bil14} and \citet{reg14} located $\sim$0.16$''$ ($\sim$19 au) north from the central star. 

In principle, both dust and ionized gas could be contributing to the radio emission at the innermost regions of transitional disks. To our knowledge, very few transitional disks have been observed with high enough angular resolution to detect and resolve compact emission at mm and cm wavelengths inside their central cavities or gaps. LkCa 15 \citep{ise14} and AB Aur \citep{rod14} were imaged with the VLA at 7 mm and 3 cm, respectively, whereas TW Hya was observed with ALMA at 0.87 mm \citep{and16}. The emission in TW Hya has been attributed to inner residual dust located close to the star, although it has also been recently suggested that a contribution of free-free emission from photoionized gas could be present \citep{erc17}. The morphology and spectral index of the emission in AB Aur indicates that it is associated with free-free emission from an accretion-driven jet. In LkCa 15, however, the lack of observations at other wavelengths makes it impossible to distinguish between a dust or a free-free origin for the emission. In the following we discuss the origin of the observed compact central radio source in HD 169142.

\begin{figure}
\figurenum{3}
\plotone{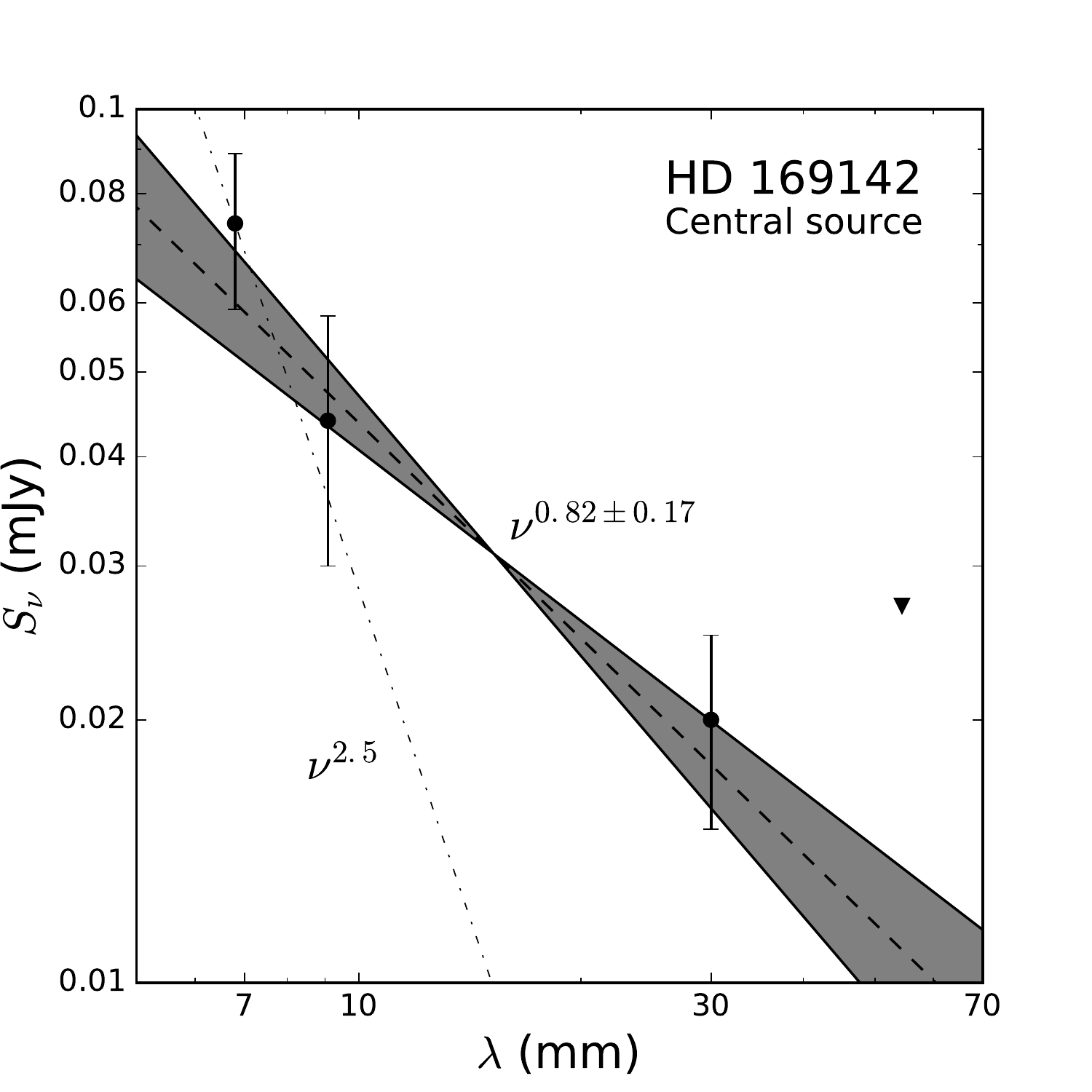}
\label{fig:SED}
\caption{Spectral energy distribution of the radio emission of the central compact radio source in HD 169142. The black points and error bars represent our measured 7 mm, 9 mm, and 3 cm flux densities. The dashed black line indicates the power-law fit to these data. The grey area represents the uncertainty (1$\sigma$) of the fit. The dot-dashed line shows, as a reference, a power-law with a spectral index of 2.5, typical of dust thermal emission. The triangle indicates the non-detection upper limit for the flux density at 5 cm (C band).}
\end{figure}

The near-IR excess in the SED of HD 169142 indicates that its disk is a pre-transitional disk, with a hot dust component located very close to the star. \citet{oso14} modeled the broadband SED as well as the 7 mm images of HD 169142 and found that an inner disk of 0.6 au in radius, together with its inner wall at the dust sublimation radius ($\sim0.2$ au), could fit the near-IR emission of HD 169142. According to their model, this inner dust component would produce only $\sim9~\mu$Jy, $\sim5~\mu$Jy and $<1~\mu$Jy of emission at 7 mm, 9 mm, and 3 cm, respectively. These values are much lower than the observed $74\pm15~\mu$Jy, $45\pm14~\mu$Jy and $\sim20\pm5~\mu$Jy at these wavelengths for the central radio source. From a power-law fit to these observed values of the flux density ($S_{\nu}\propto \nu^{\alpha}$), we estimate a spectral index $\alpha=0.82\pm0.17$ (see Fig. \ref{fig:SED}), which is too low to correspond to dust thermal emission (dust thermal emission presents $\alpha\geq2$). Therefore, both the flux density and spectral index of the central radio source indicate that its emission is  mainly dominated by partially optically-thick free-free emission from ionized gas.

Gas near young stellar objects has been found to be ionized by two main mechanisms: shocks in accretion-driven jets \citep{ang15}, and photoionization due to the high energy radiation from the central star \citep{ale14}. 
Since accretion and ejection of material are correlated \citep{cab07}, low mass-accretion rate objects, such as transitional disks, are expected to present relatively weak radio jets. However, recent studies with the VLA have shown that radio jets in this type of sources can produce free-free emission at levels that are detectable with the improved sensitivity of the VLA \citep{rod14}. 
In particular, \citet{mac16} presented VLA observations at 3 cm toward the transitional disk of GM Aur revealing resolved free-free emission from a radio jet and from a photoevaporating disk, showing that both mechanisms can contribute at the same level to the total free-free emission. 

The central radio source detected at 7 mm in HD 169142 presents a slight elongation with a PA$\simeq-80^{\circ}$, which is consistent with the position angle of the disk rotation axis (PA$\simeq-85^{\circ}$), as shown by molecular line observations \citep{ram06}. This suggests that the central radio source in our images could be tracing an accretion-driven radio jet. We can obtain a rough estimate of the free-free emission of an accretion-driven jet in HD 169142 with the empirical correlation between the radio luminosity of a source, $S_{\nu}d^2$, and its outflow momentum rate, $\dot{P}_{\rm out}$ (\citealp{ang95,ang15}):
$(\dot{P}_{\rm out}/M_{\odot}{\rm ~yr}^{-1}{\rm~km~s}^{-1})=10^{-2.5\pm0.3}(S_{\nu}d^2/{\rm mJy~kpc}^2)^{1.1\pm0.2}$.
\citet{wag15} measured a mass accretion rate onto the star HD 169142 of $\dot{M}_{\rm acc}\simeq(1.5$--$2.7)\times10^{-9}~M_{\odot}~{\rm yr}^{-1}$. Assuming a ratio between mass loss rate in jets and mass accretion rate $\dot{M}_{\rm out}/\dot{M}_{\rm acc}\simeq0.1$ \citep{cab07}, we estimate that the outflow momentum rate should be $\dot{P}_{\rm out}\simeq10^{-7}~M_{\odot}~{\rm yr^{-1}~km~s^{-1}}$. Therefore, the correlation would predict a flux density at 3 cm $S_{\nu}\simeq5~\mu$Jy, which is lower than our estimated flux density of $20\pm5~\mu$Jy for the central radio source. Even though this estimate has large  uncertainties, it suggests that an additional mechanism other than shocks associated with the outflow could contribute to the ionization of this jet. 

Another possibility is that the observed central radio source was originated as a result of photoionization by high energy radiation from the central star. Extreme-UV (EUV) and, to a lesser extent, X-rays radiation impinging on the inner disk can ionize its surface \citep{cla01,gor09,owe10} and contribute to the observed free-free emission. An inhomogeneous inner disk could lead to an inhomogeneous irradiation of its surface, which could result in the asymmetric morphology of the central radio source observed at 7 mm. Additionally, the ejected gas in the accretion-driven jet could also be significantly photoionized by the high-energy radiation emitted by the star \citep{hol09}.

Finally, the peak of emission of the central radio source could be tracing the position of an independent object at a radius of $\sim3$ au from the central star. 
Given the proximity to the central star, the dynamical timescales involved would be short. An orbiting object at a radius of $\lesssim3$ au would have an orbital period $\lesssim$4 yr, and would show detectable orbital proper motions in a few months. On the other hand, knots of emission in a radio jet are expected to be ejected with velocities of $\sim$300 km~s$^{-1}$. Given the small inclination angle of the disk ($i=13^{\circ}$, \citealp{ram06}), after projection on the plane-of-the sky, this would result in proper motions of $\sim0\rlap.''12$ yr$^{-1}$ away from the central star along PA $\simeq -80^{\circ}$. Future observations should reveal detectable variations and/or proper motions in the observed emission that will allow us to discriminate whether it traces material ejected from the central star (jet) or orbiting around it (disk or independent object).

\subsection{Outer gaps} \label{subsec:gaps}

\begin{figure}
\figurenum{4}
\plotone{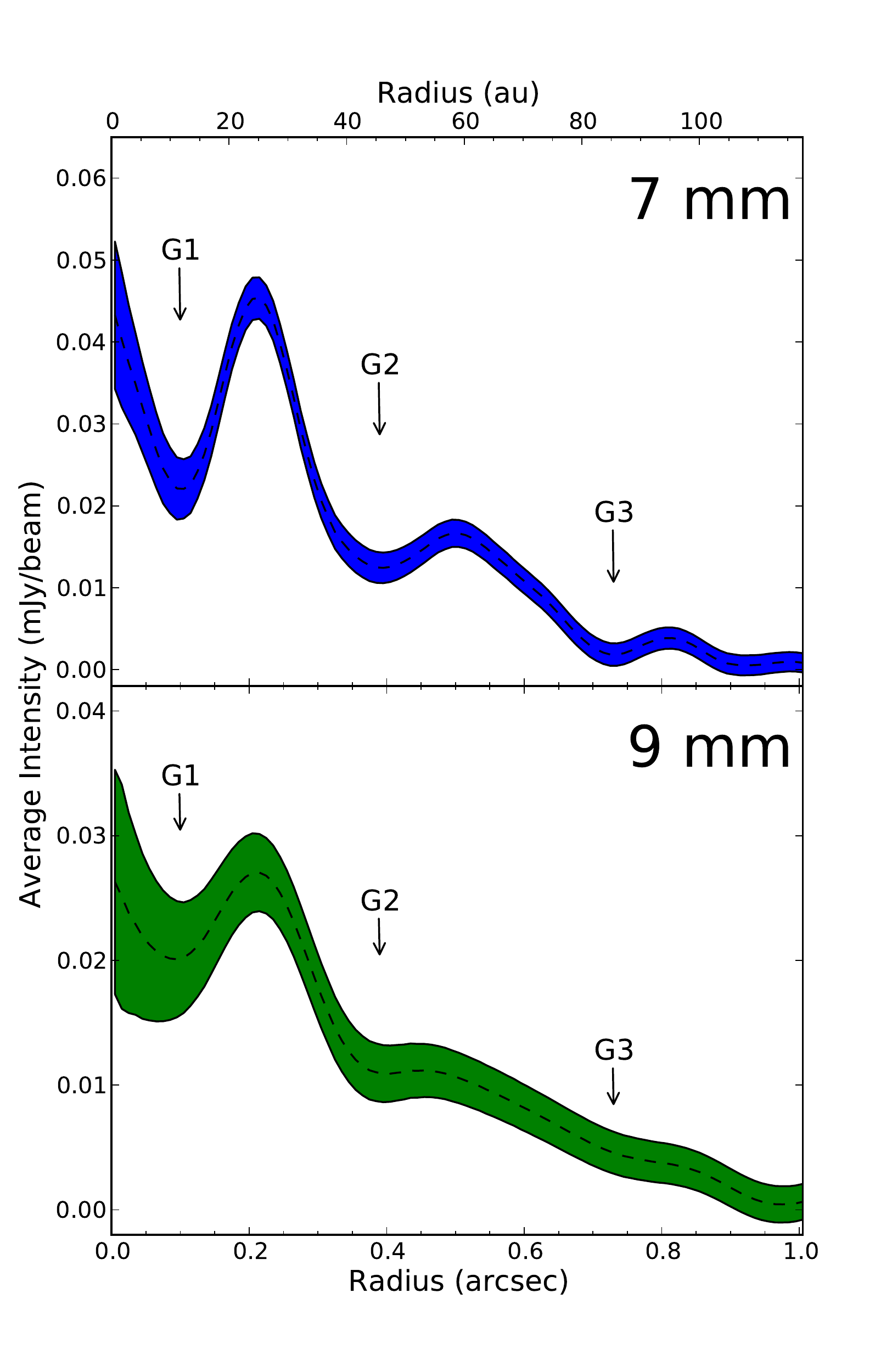}
\label{fig:profiles}
\caption{Averaged radial intensity profiles of the 7 mm (top panel) and 9 mm (bottom panel) images. The width of the lines indicates the $1\sigma$ uncertainty.}
\end{figure}

In order to improve the signal to noise ratio of the detected intensity, we have obtained the averaged radial intensity profiles of the 7 and 9 mm images (see Fig. \ref{fig:profiles}). These profiles were produced by averaging the intensity within concentric elliptical rings, matching the inclination and position angle of the disk major axis determined by previous molecular line observations ($i=13^{\circ}$, and PA$=5^{\circ}$; \citealp{ram06}). The width of the concentric rings was set to the size of the beam, since the rms noise at spatial scales smaller than a beam is not independent.  

Besides the inner gap (hereafter G1), two other gaps at radii $\sim0.38''$ ($\sim45$ au; G2) and $\sim0.73''$ ($\sim85$ au; G3) are revealed at both 7 mm and 9 mm. 
The inner (G1) and second gaps (G2) were already detected at near-IR wavelengths (with G1 at radii $\lesssim0\rlap.''19$ and G2 extending from $\sim0\rlap.''28$ to $\sim0\rlap.''48$; \citealp{qua13,mom15}), at 7 mm \citep{oso14}, and at 1.3 mm \citep{fed17}.
Our observations detect the G1 and G2 gaps at the same radii than precedent studies, while the G3 gap is reported here for the first time.

We point out that a sharp cutoff in the \textit{uv} coverage of the observations can result in the creation of a spurious annular structure during the deconvolution process. However, this artifact would appear at different radii in the images at different frequencies, whereas the G3 gap appears at the same radius in our observations at 7 and 9 mm. Additionally, our simulated images (see \S \ref{subsec:ring}) do not show the presence of spurious features mimicking the G3 gap when an actual gap is not included in the model. Therefore, we conclude that the G3 gap is probably a real annular gap in the disk of HD 169142. We note, however, that the G3 gap is not visible in the ALMA 1.3 mm images \citep{fed17}. This could indicate that the G3 gap is more prominent in the distribution of the cm-sized dust grains traced by our VLA observations.

As mentioned above, different studies have proposed that G1 and G2 are probably formed because of dynamical interactions between the disk and forming planets within each gap \citep{reg14,oso14,mom15,wag15,fed17}.
However, the origin of the third gap (G3) is more difficult to understand.
Detection of gaps at such large distances, like G3, is very difficult. So far, similar gaps have only been detected in the protoplanetary disks around HL Tau \citep{alm15}, TW Hya \citep{and16}, and HD 163296 \citep{ise16}, through recent extremely high angular resolution ALMA observations. The density in the disk mid-plane at such large distances is probably too low to create a planet responsible for clearing the observed gap. Alternatively, the magneto-rotational instability (MRI) in magnetized disks can produce pressure bumps at the outer edges of the dead zones in the disk, which can in turn trap the large dust grains and create gaps in the mm emission of the disk \citep{flo15}. However, dead zones are expected to be closer to the central star (at radii $\sim50$ au), so models do not predict gaps as far as the observed G3 gap in HD 169142. 

Another possible origin for G3 would be grain growth and/or an increase in the solids surface density close to condensation fronts (i.e. snowlines) of volatiles in the disk. This process has been suggested as the responsible for some of the rings and gaps that were detected by ALMA in the disk around the younger T Tauri star HL Tau \citep{alm15,zha15,oku16}. Models and laboratory experiments suggest that dust grains can grow significantly when surrounded by an icy mantle, which would form on the surface of the grains beyond these snowlines (\citealp{ros13,tes14} and references therein). In addition, it has been suggested that the enhanced surface density of solids beyond the snowline can produce viscosity gradients and pressure bumps because of a reduction in the column depth of the MRI-active layer. These pressure bumps could in turn trap the large dust grains \citep{kre07}, although more recent studies have found that, at least for the water snowline, which is located at much smaller radii, the viscosity gradient would need to be unrealistically high to be able to form a dust trap \citep{bit14}.

Based on the composition of comets, the most abundant volatiles in protoplanetary disks are thought to be water, CO, and CO$_2$. These molecules have, for typical disk mid-plane densities, condensation temperatures of 128--155 K, 23--28 K, and 60--72 K, respectively \citep{zha15}. Comparing these temperatures with the mid-plane temperatures obtained from the model presented by \citet{oso14}, we find that the water and CO$_2$ snowlines would fall inside G1, while the CO snowline would be located at a distance of $90$--$130$ au\footnote{We note that the \citet{oso14} model used a distance of 145 pc, which is a $\sim20\%$ larger than the recent distance of 117 pc measured by Gaia. However, we do not expect important changes in the physical structure of the disk.}.  The lower end of the range of radii for the CO snowline is coincident with the outer edge of G3, favoring grain growth and an increase in the solids surface density close to the CO snowline as a possible mechanism to explain the origin of the G3 gap.  
 
An independent measurement of the position of the CO snowline in protoplanetary disks can be obtained through observations of molecules whose chemistry is sensitive to the CO \citep{qi13}. One of these molecules is the DCO$^+$, which is expected to form mainly in the regions of protoplanetary disks where gas-phase CO and low temperatures (T$<30$ K) coexist. Because of this, the DCO$^+$ emission has been used as a tracer of the CO freeze-out in disks, showing a ring-like morphology peaking at just a slightly smaller radius than the CO snowline \citep{mat13,obe15}. However, recent studies suggest that, in some cases, the DCO$^+$ molecule might have a more complex relationship with the CO snowline, and that optically thin CO isotopologues such as C$^{18}$O could represent better tracers by showing a decrease of emission at the radius of the snowline \citep{qi15,hua17}. 

\begin{figure}
\figurenum{5}
\plotone{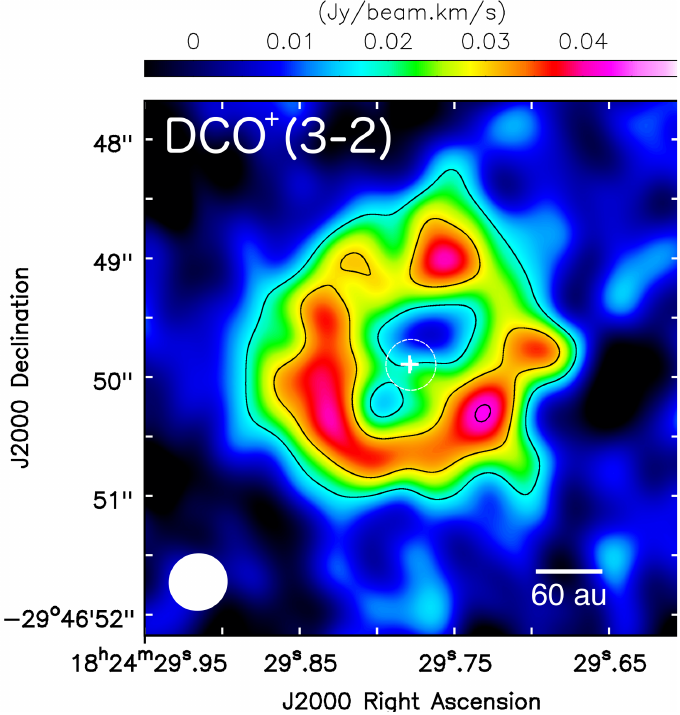}
\label{fig:dco}
\caption{ Velocity-integrated intensity (zero-order moment) of the DCO$^+$(3-2) emission of the disk around HD 169142 (synthesized beam=$0\rlap.''50\times0\rlap.''48$, PA=$-65^{\circ}$; shown in the lower-left corner). Contour levels are $-$3, 3, 5, 7, and 9 times the rms of the map, $6.0~\mu$Jy beam$^{-1}$ km s$^{-1}$. The dashed ellipse indicates our fit to the ring image at 7 mm (Fig. \ref{fig:maps}). The white cross shows the position of the star.}
\end{figure}

\begin{figure}
\figurenum{6}
\plotone{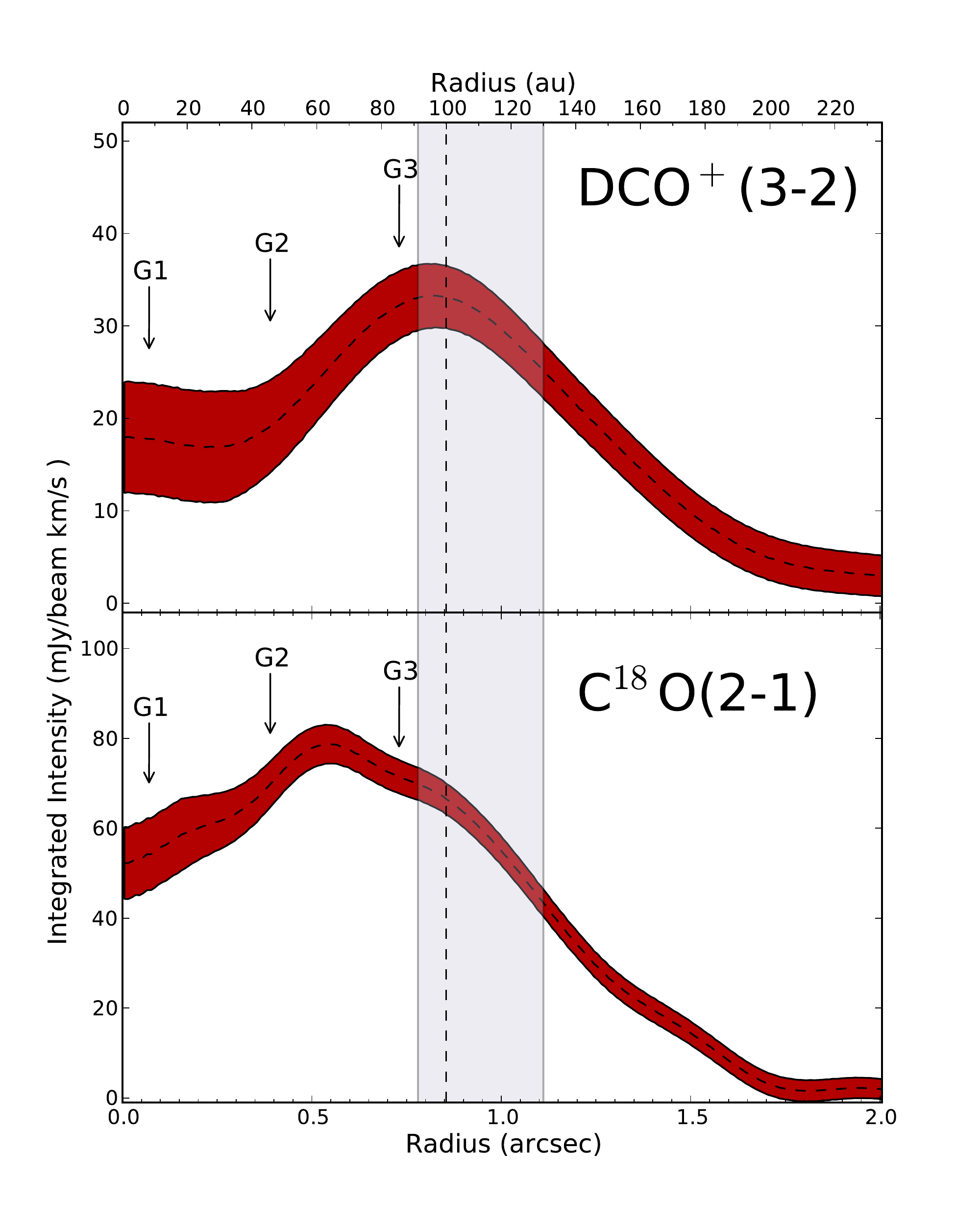}
\label{fig:gas_profiles}
\caption{Radial profiles of the averaged velocity-integrated intensity (zero-order moment) of the DCO$^+$(3-2) (top panel) and C$^{18}$O(2-1) emission (lower panel). The width of the lines indicates the $1\sigma$ uncertainty. The G1, G2, and G3 labels indicate the position of the three gaps detected at 7 and 9 mm. The shaded region between 90 and 130 au indicates the range of distances for the CO snowline estimated from the \citet{oso14} model. The vertical dashed line indicates the proposed position of the CO snowline based on the change in slope of the C$^{18}$O(2-1) emission and the position of the DCO$^+$(3-2) ring.}
\end{figure}

In order to estimate the position of the CO snowline in HD 169142, we have analyzed ALMA archival observations of both the DCO$^+$(3-2) and C$^{18}$O(2-1) molecular transitions. An image of the velocity-integrated DCO$^+$(3-2) emission is shown in Fig. \ref{fig:dco}, whereas the C$^{18}$O(2-1) images are reported by \citet{fed17}. We have obtained averaged radial profiles of the velocity-integrated emission for both transitions following the same procedures as for the continuum emission (see Fig. \ref{fig:gas_profiles}). The image of the DCO$^+$(3-2) emission shows a ring morphology (Fig. \ref{fig:dco}) with radius $\sim0\rlap.''80$ ($\sim95$ au), as it is clearly seen in the radial profile shown in Fig. \ref{fig:gas_profiles}. Additionally, the C$^{18}$O(2-1) emission shows a significant change of slope in its radial profile at $\sim0\rlap.''85$ ($\sim100$ au), indicating a decrease of the gas-phase CO beyond this radius. Thus, both molecular transitions indicate that the CO snowline is located at $\sim100$ au in the disk of HD 169142, very close to the G3 gap and in the range of distances estimated with the \citet{oso14} model.

These results suggest that the G3 gap might be produced because of an accumulation of large dust grains at the position of the CO snowline. Large grains could grow and accumulate close to the CO snowline, creating a ring of emission at long wavelengths next to an inner zone with a lower density of large grains, resulting in the observed gap in the 7 mm emission of the disk. 
As noted above, the 1.3 mm emission of the disk does not show the presence of G3 gap \citep{fed17}, which could indicate that the larger ($\sim $cm-sized) dust grains are being trapped in the CO snowline \citep{kre07}, whereas the smaller ($\sim$ submm/mm-sized) grains traced by the 1.3 mm emission have continued their inward migration.
Further studies, including both modeling and observations, are needed to fully understand the possible changes in the disk appearance with the observing wavelength.

Therefore, even though we cannot completely discard a magnetically induced or planet-induced origin, we favor dust growth and accumulation of large dust grains close to the CO snowline as the most likely mechanism responsible for the proposed G3 gap in HD 169142. 

\section{Summary and conclusions} \label{sec:conclusions}

We have presented new high angular resolution VLA observations at 7 mm, 9 mm, and 3 cm toward the pre-transitional disk around HD 169142. Our main results can be summarized as follows:

\begin{itemize}
\item Our 7 and 9 mm observations show a narrow ($\sim8$ au in width) azimuthally asymmetric ring of emission of radius $\sim25$ au, apparently tracing the outer rim of the innermost gap. The radius of the ring is consistent with that of previous 7 mm and near-IR polarized scattered light images \citep{qua13,oso14,mom15}. A bright knot of emission in the ring is revealed at 7 mm. This knot  of emission is not present in the near-IR images, indicating that it is probably tracing an azimuthal asymmetry in the density of the disk mid-plane, to which our 7 mm images are more sensitive. We interpret this asymmetry as probably produced by tidal interactions between the disk and forming planets. 

\item A central component of emission is detected inside the inner gap at 7 mm, 9 mm, and 3 cm. The 7 mm source shows a slightly elongated morphology approximately along the E-W direction, with its peak of emission displaced a projected distance of $\sim3$ au ($\sim0\rlap.''025$) to the west of the central star.  
The flux density and spectral index of this central radio source indicate that it is dominated by free-free emission from ionized gas, which could be associated with an inhomogeneous photoionization of the inner disk, with an independent orbiting object, or with an (asymmetric) ionized jet. Although data currently available seem to favor the latter scenario, future observations should reveal significant proper motions either away or around the central star that will allow us to discriminate between ejection or orbital motions.

\item The radial intensity profiles of the 7 and 9 mm images reveal the presence of multiple gaps in the disk of HD 169142. Our 7 and 9 mm observations not only confirm the presence of the previously reported inner (G1) and second (G2) gaps, which approximately extend through radii $\sim$0--25 au and $\sim$32--56 au, respectively, but also detect, for the first time, a new gap comprising the radii $\sim$77--96 au (G3). This proposed gap, one of the farthest gaps ever detected in a protoplanetary disk, is not detected in ALMA 1.3 mm images, suggesting that it might be more prominent in the distribution of the larger dust grains traced by our VLA observations.

\item Our analysis of DCO$^+$(3-2) and C$^{18}$O(2-1) ALMA observations, as well as the results of the \citet{oso14} model, indicate that the CO snowline is located at $\sim100$ au. This suggests that dust grain growth and an increase in the solids surface density close to the CO snowline could be the mechanism responsible for the origin of the proposed G3 outer gap.

\end{itemize}

\acknowledgments

We thank an anonymous referee for useful and valuable comments. This paper makes use of the following ALMA data: ADS/JAO.ALMA\#2013.1.00592.S. ALMA is a partnership of ESO (representing its member states), NSF (USA) and NINS (Japan), together with NRC (Canada), NSC and ASIAA (Taiwan), and KASI (Republic of Korea), in cooperation with the Republic of Chile. The Joint ALMA Observatory is operated by ESO, AUI/NRAO and NAOJ. E.M., G.A., M.O., J.M.T, and J.F.G. acknowledge support from MINECO (Spain) grant AYA2014-57369-C3 (co-funded with FEDER funds). JMT acknowledges support from the Generalitat de Catalunya/CERCA programme. C.C.-G. acknowledges support from UNAM-DGAPA PAPIIT IA102816.

\vspace{5mm}
\facilities{VLA}

\software{CASA (v 4.3.1, 4.5.3; \citealp{mcm07})}

\clearpage

\listofchanges

\end{document}